%% file: entropy.tex
\documentclass[aps,prl,showpacs,groupedaddress]{revtex4}  
\usepackage{graphicx}  
\usepackage{dcolumn}   
\usepackage{bm}        
\usepackage{amssymb,amsmath}   
\usepackage{amsthm}
\usepackage{subfigure}

\newtheorem{theorem}{Theorem}[section]

\newtheorem{example}[theorem]{Example}

\begin{document}



\title{Symmetry based Structure Entropy of Complex Networks}
\input list_of_authors_r2.tex  
\date{\today}

\begin{abstract}
Precisely quantifying the heterogeneity or disorder of a network
system is very important and desired in studies of behavior and
function of the network system. Although many degree-based entropies
have been proposed to measure the heterogeneity of real networks,
heterogeneity implicated in the structure of networks can not be
precisely quantified yet. Hence, we propose a new structure entropy
based on automorphism partition to precisely quantify the structural
heterogeneity of networks. Analysis of extreme cases shows that
entropy based on automorphism partition can quantify the structural
heterogeneity of networks more precisely than degree-based
entropies. We also summarized symmetry and heterogeneity statistics
of many real networks, finding that real networks are indeed more
heterogenous in the view of automorphism partition than what have
been depicted under the measurement of degree-based entropies; and
that structural heterogeneity is strongly negatively correlated to
symmetry of real networks.
\end{abstract}

\maketitle

\section{\label{sec:level1}Introduction}
In recent years, great efforts have been dedicated to the research
on complex networks, due to the fact that many complex systems can
be modeled as networks consisting of components as well as relations
among these components. Previous studies primarily focus on finding
various statistical properties of real networks, especially degree
based statistics, such as degree
distribution\cite{barabasiBA,Albert}, degree
correlation\cite{RPastor,Maslov,jberg}, degree-based structure
entropies\cite{wangb,sole}. Studies of many significant properties
of networks, such as heterogeneity\cite{barabasiBA}, assortative
mixing\cite{Newman1,Newman2} and self-similarity \cite{song,zhang},
are based on these statistics.

Degree delivers to us the most important information about the
number of interconnections of each individual component in the
network. However, degree only provides us a view of complex networks
in a shallow level, for the reason that vertex partition \footnote{
A \emph{partition} of the set $V$ is a set of disjoint non-empty
subsets of $V$ whose union is $V$. Elements of a partition are also
called its \emph{cells}. A \emph{trivial cell} is the cell with
cardinality one. If every cell of a partition is trivial, then the
partition is a \emph{discrete partition}; while if the partition
only has one cell, the partition is a \emph{unit partition}. For two
partitions on set $V$, $P$ and $Q$, if every cell of $P$ is a subset
of some cell of $Q$, we say that $P$ is finer than $Q$, or $Q$ is
coarser than $P$. } based on degree is coarser than many finer
vertex partitions in many networks, e.g., automorphism partition ---
a core concept in the symmetry of network. In other words, in some
networks, vertex with the same degree would be further
differentiated from each other, thus forming a finer partition.
Consequently, a fascinating problem arises, what will complex
network looks like if automorphism partition is employed instead of
degree partition? Since degree-based statistics are the driving
forces of many existing studies in complex networks. We believe that
studies of complex network in the view of symmetry will open a brand
new field leading us to deeper understanding about complex networks.

 There is increasingly recognition that measuring heterogeneity of complex networks is very important
  in studies of behavior and function of complex networks.
 It has been shown in \cite{Albert} that heterogeneity of degree is
 directly related to the robust-yet-fragile property of scale-free networks ,
i.e., robustness against random failures of vertices but
vulnerability to target attacks. Furthermore, it has been found in
\cite{nish} that the homogeneous networks are more synchronizable
than heterogeneous ones, even though the average network distance is
large.

However, existing heterogeneity measures \cite{sole,wangb} of
complex networks are all based on degree. Specifically, entropy in
\cite{sole} is based on remaining degree \cite{Newman1,Newman2}
distribution and entropy in \cite{wangb} is based on degree
distribution. In fact, degree-based measures of heterogeneity are
only the precise quantification of degree heterogeneity of networks,
not that of actual heterogeneity in the sense of structure of many
networks. To some extent, degree heterogeneity of networks is only
the approximations of structural heterogeneity of networks. For
example, as shown in Example \ref{exa:cube}, in some networks,
vertices with the same degree still can be differentiated from each
other through measurement on some structural properties of
individual vertex, such as the number of triangles passing through a
vertex, the shortest path passing through a vertex(also know as
betweenness \cite{Freeman}). Hence, heterogeneity measured by degree
partition can not precisely describe the structural heterogeneity
for all networks. Luckily, automorphism partition of the network
naturally partitions the vertex set into structurally equivalent
cells, thus offering us an ideal alternative to measure the
heterogeneity of network structure.

\begin{figure}
\centering
 \subfigure[cuneane] { \label{fig:regular:a}
\includegraphics[scale=0.8]{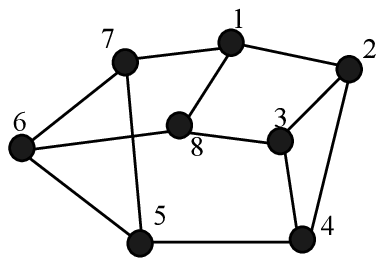}}
\subfigure[$Q_3$] { \label{fig:regular:b}
\includegraphics[scale=0.65]{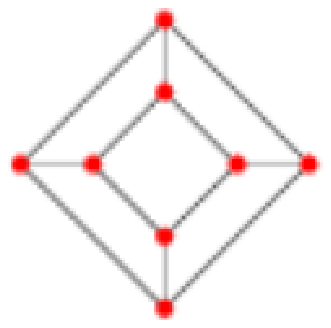}}
\caption{Illustration of two 3-regular graphs. Figure (a) shows an
abstract structure of molecule known as \emph{'cuneane'}, which is
not vertex transitive. Figure (b) shows an example of 3-cube graph,
denoted as $Q_3$, which is vertex transitive.} \label{fig:regular}
\end{figure}

\begin{example}
As shown in Figure \ref{fig:regular}, \emph{'cuneane'} and $Q_3$ are
all regular graphs with degree 3. Hence, structure of
\emph{'cuneane'} and  $Q_3$ can be considered as completely
homogeneous if degree-based entropy measures are utilized. However,
intuitively, we can see that homogeneity in \emph{'cuneane'}is
different from that of $Q_3$. All the vertices in $Q_3$ are
equivalent from the structure perspective, thus forming a \emph{unit
partition}, and we can not further partition the vertex set.
However, in \emph{'cuneane'}, we can easily find that vertices 1 and
8 play a different role from that of vertices 4 and 5 or the
remaining vertices, for the reason that 1 and 8 are the only ones
not involved in the two triangles, 4 and 5 are the only ones
connected by an edge between two different triangles. Therefore, for
\emph{'cuneane'}, we can construct a vertex partition
$\mathcal{P}=\{\{1,8\},\{4,5\},\{2,3,6,7\}\}$, which is finer than
degree partition. Furthermore, we can validate that partition
$\mathcal{P}$ is just the automorphism partition of
\emph{'cuneane'}. \label{exa:cube}
\end{example}

\section{\label{sec:level1}Symmetry-based Structure Entropy}

A graph is denoted as $G=G(V,E)$, where $V$ is the set of vertices
and $E\subseteq V\times V$ is the set of edges. If $(v_1,v_2)\in E$,
then we say that $v_1$ and $v_2$ are adjacent. An automorphism
acting on the vertex set can be viewed as a permutation of the nodes
of the graph preserving the adjacency of the vertices. The set of
automorphisms under the product of permutation forms a
group\cite{Godsil}. In general, a network is \emph{asymmetric} if
its automorphism group is the identity group, which only contains a
identity permutation; otherwise, the network is \emph{symmetric}. A
graph $G=G(V,E)$ is \emph{vertex transitive}(or just transitive) if
its automorphism group acts transitively on $V$, which means that
for any two distinct vertices of $V$, there is an automorphism
mapping one to the other.

Let $Aut(G)$ be the automorphism group acting on vertex set $V$.
Then naturally, we can get a partition
$\mathcal{P}=\{V_1,V_2,...,V_k\}$, called as automorphism partition,
 in the way that $x$ is equivalent to $y$ if and only if $\exists
g\in Aut(G)$, s.t. $x^g=y$. And each cell of the partition is called
as an orbit of the automorphism group $Aut(G)$. Automorphism
partition offers us an in-depth insight into the heterogeneity of
networks. Compared to the degree partition of the vertex set,
automorphism partition is much finer than the degree partition for
most of networks.

To accurately measure the structural heterogeneity of complex
networks, we define an \emph{entropy based on automorphism
partition}, abbreviated as EAP, as follows:
\begin{equation}
EAP=-\sum_{1\leq i \leq |\mathcal{P}|}{p_i}\log{p_i}
 \label{equ:eap}
\end{equation},
where $\mathcal{P}$ is the automorphism partition of the network,
$p_i$ is the probability that a vertex belongs to the cell $V_i$ of
$\mathcal{P}$.  Note that given a network's automorphism partition
$\mathcal{P}=\{V_1,V_2,...,V_k\}$, we can calculate $p_i$ as:
\begin{equation}
p_i=\frac{|V_i|}{\sum_j|V_j|}=\frac{|V_i|}{N}
 \label{equ:pi}
\end{equation}, where $N$ is the number of vertices in a graph.

\begin{figure}
\centering
\includegraphics[scale=0.7]{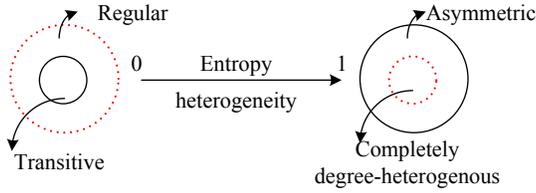}
\caption{Illustration of extreme cases under different entropy
measures. The dotted circles represent the two extreme cases of
NEDD. The solid circles represent the two extreme cases of NEAP. The
embedding relation between circles express the containment relation
between network set. Note that, the minimal cases of NERDD also can
be represented by the left dotted circle, while the maximal case of
NERDD should lie in the middle range of [0,1] in terms of the
measurement of NEDD or NEAP. } \label{fig:extreme}
\end{figure}

Obviously, the maximum value of EAP or $EAP_{max}$ equals to
$log(N)$, obtained when $p_i=\frac{1}{N}$ for each $1\leq i \leq
|\mathcal{P}|$, i.e., the graph has a \emph{discrete} automorphism
partition. The minimum value of EAP or $EAP_{min}$ equals to $0$ and
occurs when the automorphism partition is a \emph{unit partition},
implying that all the vertex belong to the same cell or all vertex
are equivalent in the structure of the network. The maximum value of
EAP corresponds to the completely structure-heterogeneous network,
i.e. asymmetric network, and the minimum value of EAP corresponds to
the completely structure-homogeneous network, i.e. transitive
networks(shown in Figure \ref{fig:extreme}) .

The normalized entropy based on automorphism partition (NEAP) can be
defined as:
\begin{equation}
NEAP=\frac{EAP-EAP_{min}}{EAP_{max}-EAP_{min}}=\frac{EAP}{\log N}
 \label{equ:neap}
\end{equation}, where $N$ is the number of vertices in the network.

For comparison, we denote entropy based on remaining degree
distribution by ERDD \cite{sole} and entropy based on degree
distribution by EDD \cite{wangb}. We also define their corresponding
normalized entropy similar to Equation \ref{equ:neap}, which are
denoted by NERDD and NEDD respectively. Example
\ref{exp:entropy_comp} illustrates the computation of these three
entropies.

\begin{example}
As shown in Figure \ref{fig:regular}, since \emph{'cuneane'} is a
regular graph, we have $EDD=ERDD=NEDD=NERDD=0$. However, the
automorphism partition of \emph{'cuneane'} is not a \emph{unit
partition}, and we have $p_1=p_2=\frac{1}{4}$, $p_3=\frac{1}{2}$.
Thus
$EAP=-\frac{1}{4}\log\frac{1}{4}-\frac{1}{4}\log\frac{1}{4}-\frac{1}{2}\log\frac{1}{2}=\frac{1}{2}\log8$,
$NEAP=\frac{\frac{1}{2}\log8}{\log8}=0.5$,
 which is a value larger than 0, thus could quantify \emph{'cuneane'} as
 heterogenous in a certain degree rather than completely homogenous.
Hence, in this case, EAP is more appropriate for quantifying
structure-heterogeneity than ERDD and EDD. \label{exp:entropy_comp}
\end{example}

  The maximum values of ERDD or EDD are both $\log(N)$,
however, the maximal values of two entropies correspond to two
different kinds of networks, respectively. For EDD, the maximal
entropy value corresponds to the completely degree-heterogenous
networks, i.e., networks with $N$ nodes partitioned into $N$
non-empty cells. For ERDD, the maximal entropy value corresponds to
the completely remaining-degree-heterogenous networks, i.e.,
networks with remaining degree equally distributed.

\begin{figure}
\centering
\includegraphics[scale=1]{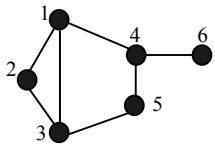}
\caption{Illustration of an asymmetric graph. The degree partition
$\mathcal{D}=\{\{1,3,4\},\{2,5\},\{6\}\}$ is much coarser than
automorphism partition, which is a discrete partition in this graph.
In the cell \{1,3,4\} of degree partition, all vertices have degree
3, however, vertex 4 is the only one adjacent to a vertex with
degree 1, which could distinguish vertex 4 from \{1,3,4\}. Vertex 1
is adjacent to two vertices with degree 3 , while vertex 3 is only
adjacent to one vertex with degree 3, which could differentiate
vertex 1 from vertex 3. Hence, vertices 1,3,4 are not
structure-equivalent to each other. Vertex 2 and 5 in the cell
\{2,5\} of degree partition also can be differentiated from each
other, because adjacent nodes of vertex 2 and adjacent nodes of
vertex 5 are not structural equivalent, i.e. vertex 1 and vertex 4
are not structural equivalent.} \label{fig:asy}
\end{figure}

Completely degree-heterogenous networks are the most heterogenous
cases under entropy measure of EDD. However, as shown in Figure
\ref{fig:asy}, completely structure-heterogenous networks are not
necessarily completely degree-heterogenous, note that the inverse
statement necessarily holds true. As long as a network is
asymmetric, i.e., the automorphism group contains no non-trivial
permutations, the network structure will be completely heterogenous.
Hence, extreme heterogenous cases should be extended to asymmetric
networks provided that more precise evaluation of structural
heterogeneity is desired (shown in Figure \ref{fig:extreme}).

The minimum values of ERDD and EDD both equal to 0, both
corresponding to regular networks, which are the most homogeneous
networks under these two entropy measures. However, as shown in
Example \ref{fig:regular}, regular graphs can be subdivided into
transitive and non-transitive graphs, and only transitive graphs are
the extreme structure homogeneous networks. Hence, extreme
homogeneous cases should be limited to transitive networks if more
precise evaluating of structural heterogeneity is desired (shown in
Figure \ref{fig:extreme}).

  According to the above facts, the relation between
degree-based entropies and symmetry based entropy also can be stated
as following statements:
\begin{enumerate}
    \item   $f_{EDD}(G)=EDD_{max}\Rightarrow
f_{EAP}(G)=EAP_{max}$, however, it does not necessarily hold true
vice versa;
    \item  $f_{EAP}(G)=EAP_{min}\Rightarrow
f_{EDD}(G)=EDD_{min}$, however, it does not necessarily hold true
vice versa;
    \item $f_{EDD}(G)=EDD_{min}\Leftrightarrow
    f_{ERDD}(G)=ERDD_{min}$;
\end{enumerate}
where $f_{EDD}$, $f_{ERDD}$ and $f_{EAP}$ are the functions
obtaining EDD, ERDD and EAP for each graph, respectively.

\begin{table*}
\caption{\label{tab:entropy} Statistics of some real networks and
theoretic networks. Summarized statistics include some basic
information about the network (All the networks are preprocessed as
an undirected, unweighted graphs without any self-loops and
multi-edges) including the number of the nodes $N$, the number of
the edges $M$, the average degree $z$. The key measures quantifying
symmetry of the network are also summarized, including the
automorphism group size of the real networks
$\alpha_G$\cite{bollobasMGT} (to simplify the representation, we use
$\lg\alpha_G$); the ratio of $\alpha_G$ to the maximal automorphism
group size of graphs with $N$ nodes, defined as
$\beta_G=(\alpha_G/N!)^{1/N}$\cite{sym1,sym2}; the ratio of number
of nodes in the non-trivial orbits to $N$, defined as
$\gamma_G=\frac{\sum_{{1\leq i\leq k,|V_i|>1}}|V_i|}{N}$\cite{xiao}.
We also generate four BA \cite{barabasiBA} networks with $m$ (the
number of nodes that a new node attach to)varying from 1 to 4 in
increment of 1. And we generate four ER\cite{ER} networks with
average degree approximately as one of \{2,4,6,8\}, using
PAJEK\cite{pajek}.}
\begin{ruledtabular}
\begin{tabular}{cccccccccc}
Network&N&M&z&$\lg\alpha_G$&$\beta_G$&$\gamma_G(\%)$&NEDD&NERDD&NEAP
\\ \hline

Technique Network\\
USPowerGrid\cite{uspowergrid}&4942&6594&2.67&152.71&$5.90\times10^{-4}$&16.7&0.20&0.25&0.98\\
InternetAS\protect\footnote{Here, the snapshot at
2006-07-10 of CAIDA\cite{caida} is used}&22443&45550&4.06&11346&$3.8784\times10^{-4}$&76.1&0.16&0.39&0.84\\

\hline

Social Network\\
arXiv\protect\footnote{Here, the snapshot at 2003-04 of HEP每TH
(high energy physics theory) citation graph \cite{arXiv} is used}
&27771&352285&25.37&333.26&$1.01\times10^{-4}$&3.51&0.41&0.51&0.99\\
USAir97\cite{USAir}& 332 & 2126 & 12.81& 24.41& $9.59\times10^{-3}$&
26.20
& 0.539 &0.68&  0.95\\
PairsP\cite{wordassc} & 10617  & 63782  & 12.02 & 632.80 &
$2.90\times10^{-4}$&24.32 & 0.32 & 0.47& 0.97\\
foldoc\cite{foldoc}&13356&91471&13.6974&17 & $2\times10^{-4}$&0.80 &
0.32 & 0.39
& 1\\
Erdos02\cite{erdos}&6927&11850&3.42&4222.5&$1.6\times10^{-3}$&73.75&0.15&
0.44&0.77\\

\hline

Biological Network\\
BioGrid-SAC\cite{biogrid}&5438&73054&13.43&57.79&$5.12\times 10^{-4}$&3.2739&0.48&0.61&1.00\\
BioGrid-MUS\cite{biogrid}&219&400&3.65&126.93&$4.69\times10^{-2}$&77.98&0.28&0.47&0.64\\
BioGrid-HOM\cite{biogrid}&7523&20029&5.32&935.09&$4.81\times10^{-4}$&24.47&0.28&0.43&0.94\\
BioGrid-DRO\cite{biogrid}&7529&25196&6.69&624.32&$4.27\times10^{-4}$&21.36&0.30&0.45&0.96\\
BioGrid-CAE\cite{biogrid}&2781&4350&3.13&829.69&$1.94\times10^{-3}$&51.08&0.21&0.411&0.85\\
ppi\cite{ppi}&1870&2203&4.7123&518.6&$2.7\times10^{-3}$&53.32 &0.21
& 0.34
&0.82\\

\hline

Theoretic Networks\\
Star Graph&2000&1999&1.99&5732.2&0.9962&99.95&$5.65\times10^{-4}$&0.09&$5.65\times10^{-4}$\\
BA(1)&2010&2000&1.99&282.09&$1.90\times10^{-3}$&56.37&0.17&0.30&0.91\\
BA(2)&2010&4000&3.98&0.60&$1.40\times10^{-3}$&0.2&0.24&0.35&1\\
BA(3)&2010&6000&5.97&0   &$1.35\times10^{-3}$& 0 &0.28 &0.39&1\\
BA(4)&2010&8000&7.96&0 &$1.35\times10^{-3}$&  0 &0.31& 0.43& 1\\
ER(1)&2000&2081&2.08&507.97&$2.4\times10^{-3}$&34&0.225&0.228&0.89\\
ER(2)&2000&4002&4&51.33&$1.4\times10^{-3}$&2.65&0.276&0.274&0.99\\
ER(3)&2000&5923&5.90& 2.07&$1.36\times10^{-3}$& 0.25 &0.30 & 0.30&1\\
ER(4)&2000&8137&8.14& 0 &$1.36\times10^{-3}$&  0 &0.32& 0.32& 1\\

\end{tabular}
\end{ruledtabular}
\label{tab:slp}
\end{table*}

\section{\label{sec:level1}Analysis}
\begin{figure}
\centering
 \subfigure[NEDD distribution] { \label{fig:hist:a}
\includegraphics[scale=0.4]{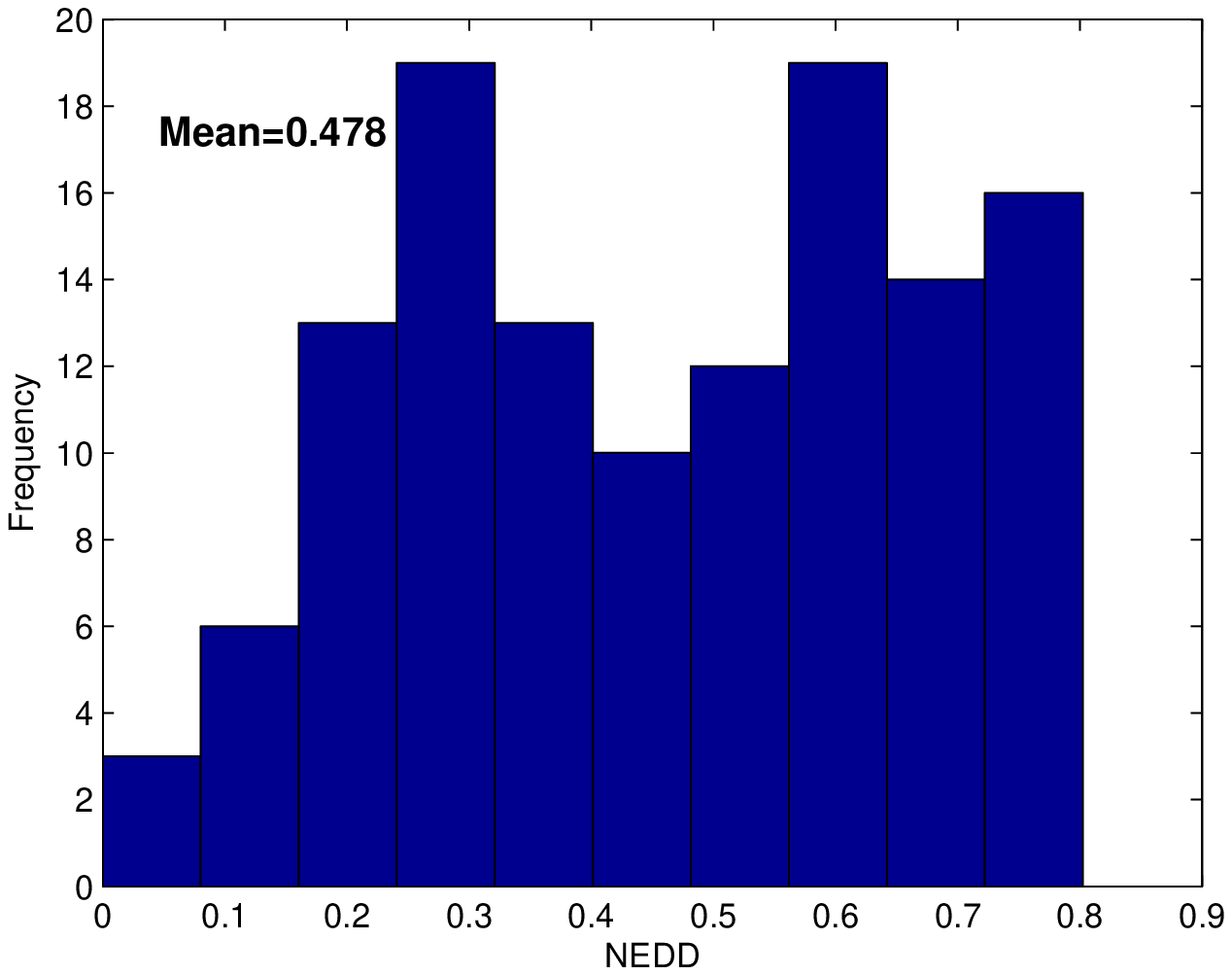}}
\subfigure[NERDD distribution] { \label{fig:hist:b}
\includegraphics[scale=0.4]{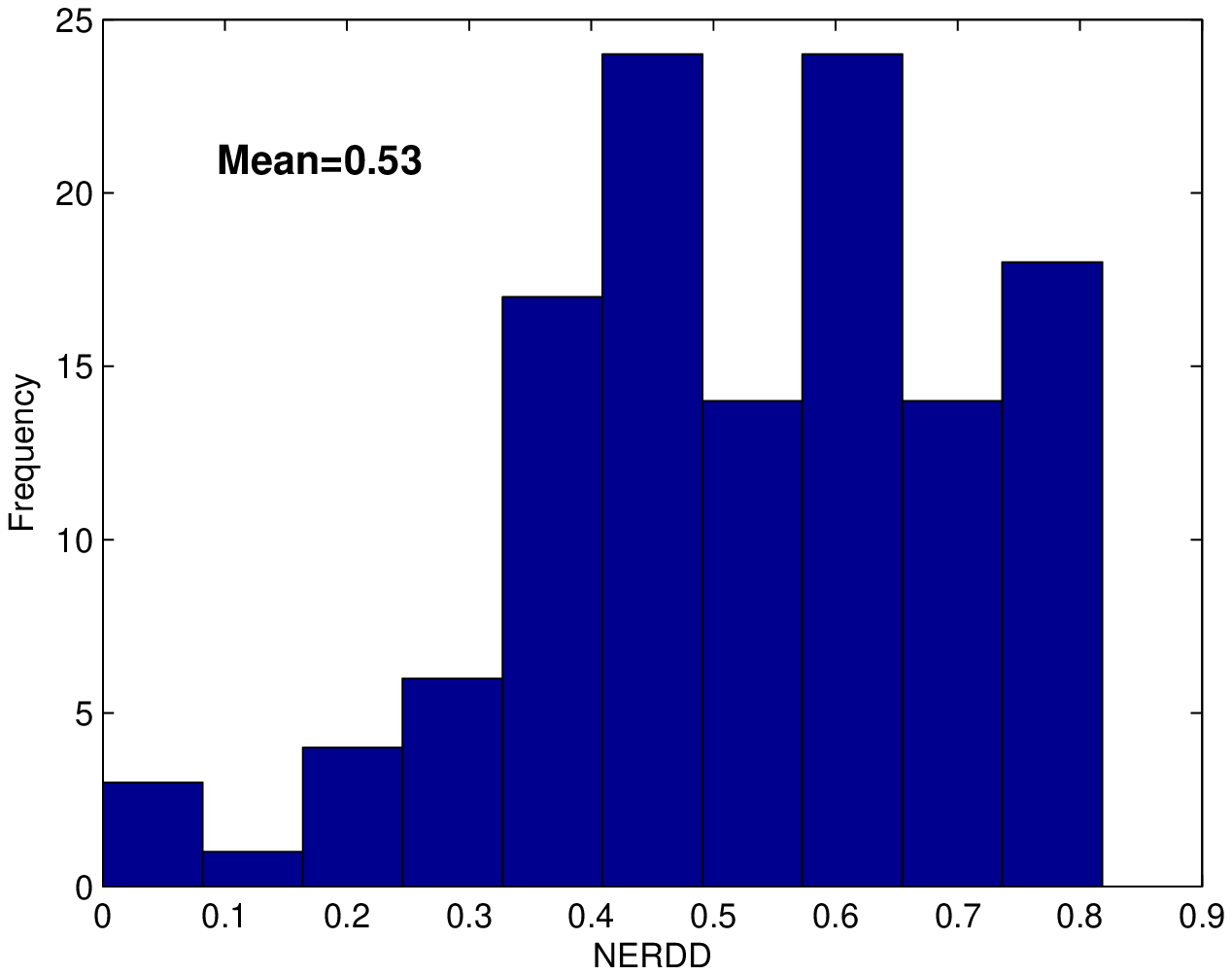}}
\subfigure[NEAP distribution] { \label{fig:hist:b}
\includegraphics[scale=0.4]{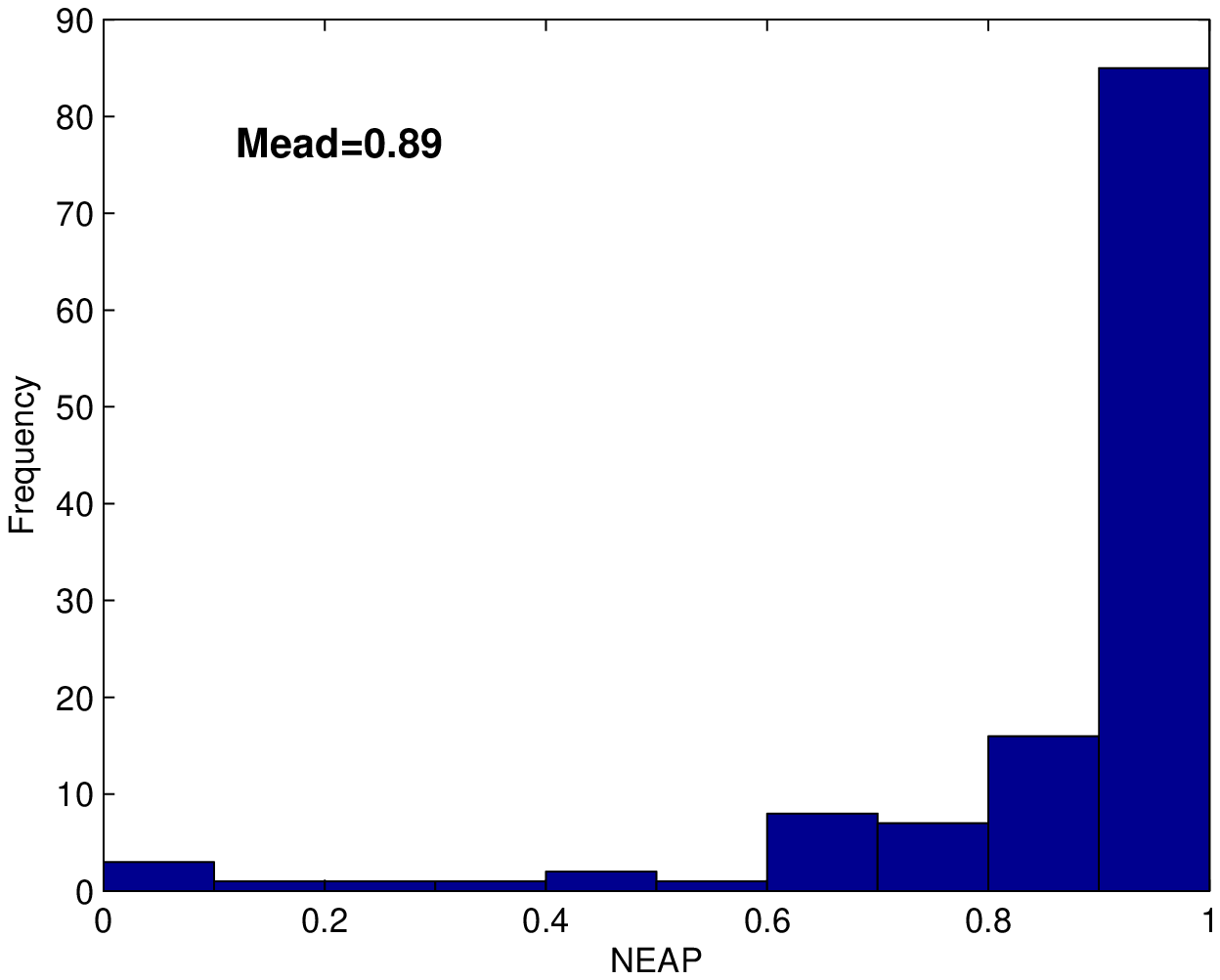}}
\caption{Distribution of values of three entropy measures, NEDD,
NERDD and NEAP for 125 real networks.} \label{fig:hist}
\end{figure}

\begin{figure}
\centering
\includegraphics[scale=0.4]{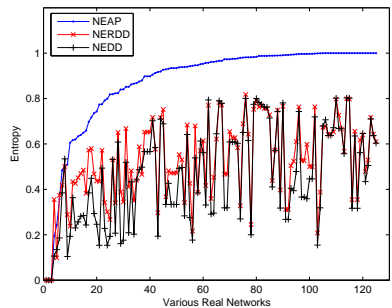}
\caption{Comparison of three entropy measures. The horizontal axis
represent various networks in the ascending order of corresponding
NEAP values. } \label{fig:comp}
\end{figure}

In this section, we first show that in the view of symmetry, or
automorphism partition, most of real networks should be
characterized as more heterogenous than what have been shown in the
view of degree partition. To show this, we calculated NEDD, NERDD
and NEAP for 125 real networks. As shown in Figure \ref{fig:hist},
NEDD values of real networks tend to lie in the range [0.2,0.8]
(overall $87.2\%$ networks lie in this range) and its mean value is
0.47; NERDD values of real networks tend to lie in the range
[0.4,0.8] (overall $75.2\%$ real networks lie in this range) and its
mean value is 0.53; while NEAP of real networks primarily lies in
the range [0.8,1] (overall $80.8\%$ real networks lie in this range)
and its mean value is 0.89, close to 1. In addition, for almost all
the tested real networks, the value of NEAP is larger than that of
NEDD and NERDD, which is shown in Figure \ref{fig:comp}. Hence, from
these observations, we can see that real networks are very
heterogeneous in the view of automorphism partition, and real
networks will have a larger probability (larger than 80\% in our
samples) to be quantified with a NEAP value larger than 0.8.

We also need to note that some real networks characterized as very
homogenous in the view of degree partition have been quantified as
very heterogenous in the view of symmetry. As shown in Table
\ref{tab:entropy}, for almost all the real networks, the
corresponding values of the degree-based entropies are less than 0.5
, except for the NERDD of arXiv, USAir97, BioGrid-SAC, and NEDD of
USAir97; while for all the networks, the corresponding values of
NEAP are larger than 0.6 and most of them larger than 0.8. If the
median value of range [0,1] is taken as the critical value
indicating whether a network is heterogenous, then many real
networks under the measure of degree based entropies, are all tend
to be quantified as homogenous. On the contrary, real networks under
the measure of symmetry-based entropy tend to be quantified as very
heterogenous.

 Since heterogeneity based on
automorphism partition describes the structure heterogeneity more
accurately than degree-based heterogeneity, it's reasonable to
believe that most of real networks are very heterogenous in their
structure.

In Table \ref{tab:entropy}, statistics of some theoretic networks
are also summarized. We can see that NEAP of star graph is close to
0, indicating that star graph is very homogenous. Indeed, in a star
graph, almost all nodes except for the central node lie in same cell
of automorphism partition and these nodes can not be differentiated
from each other by any means, thus it's natural that star graph is
very homogenous.

Generally, scale free networks are considered to be more
heterogenous than ER random networks \cite{barabasiBA}, which is
based on the intuitive observation that scale free networks are
right-skewed in double-log degree distribution while the degrees of
ER random networks are exponentially distributed with an obvious
scale. However, no quantification or theoretic proof has been
provided to verify the above notion, which can be partly attributed
to the lack of appropriate measures of heterogeneity of real
networks.  However, utilizing three entropy measures, we can clearly
see that under the measurement of NEDD and NERDD, the difference of
heterogeneity between scale free and random networks are very
small(less than 0.05), and under the measure of NEAP, both scale
free and random networks tend to be quantified as very heterogenous
without clear difference (less than 0.02).

\begin{figure}
\centering
 \subfigure[Correlation between $NEAP$ and $\beta_G$] { \label{fig:correlation:a}
\includegraphics[scale=0.4]{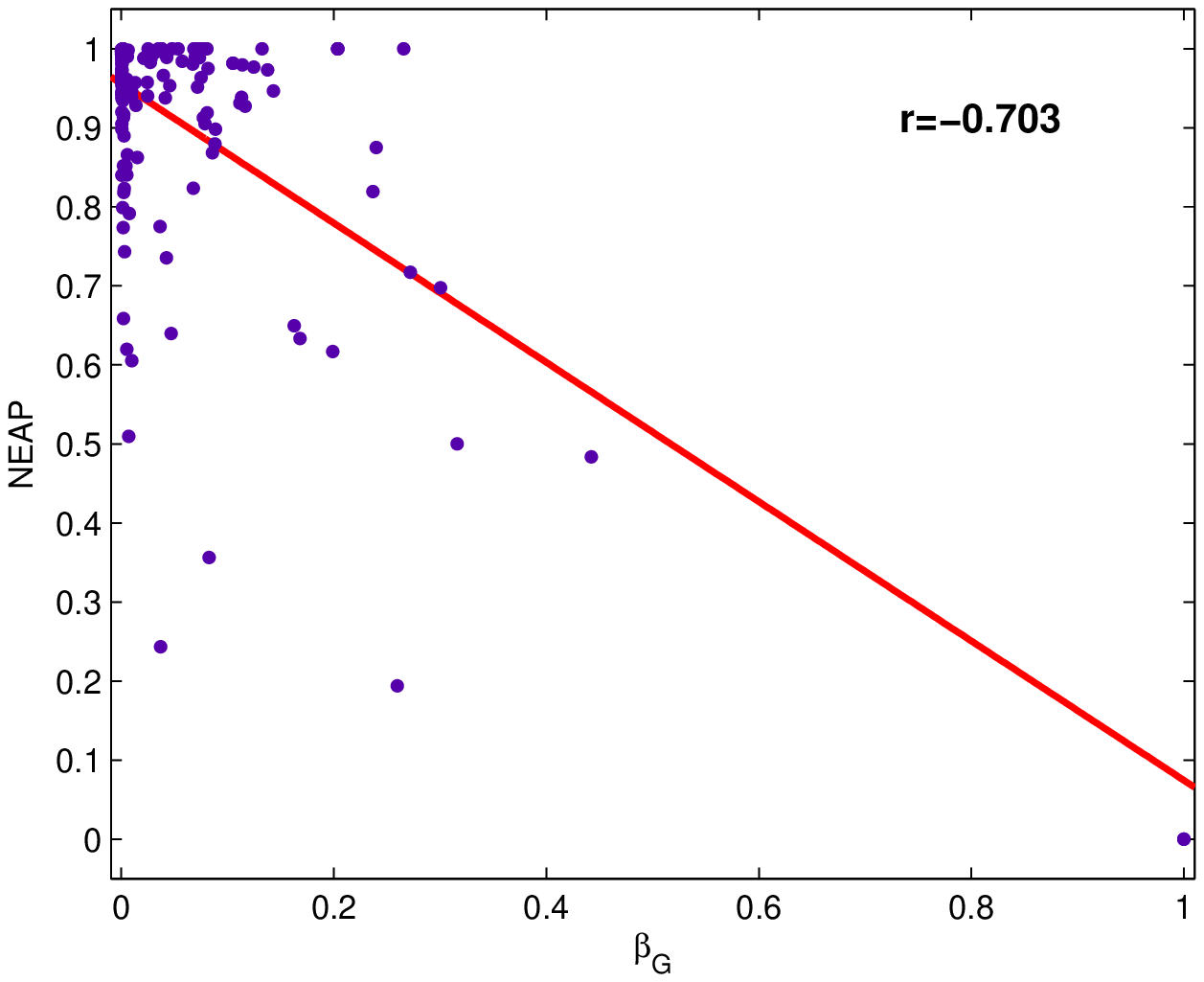}}
 \subfigure[Correlation between $NEAP$ and $\gamma_G$] { \label{fig:correlation:b}
\includegraphics[scale=0.4]{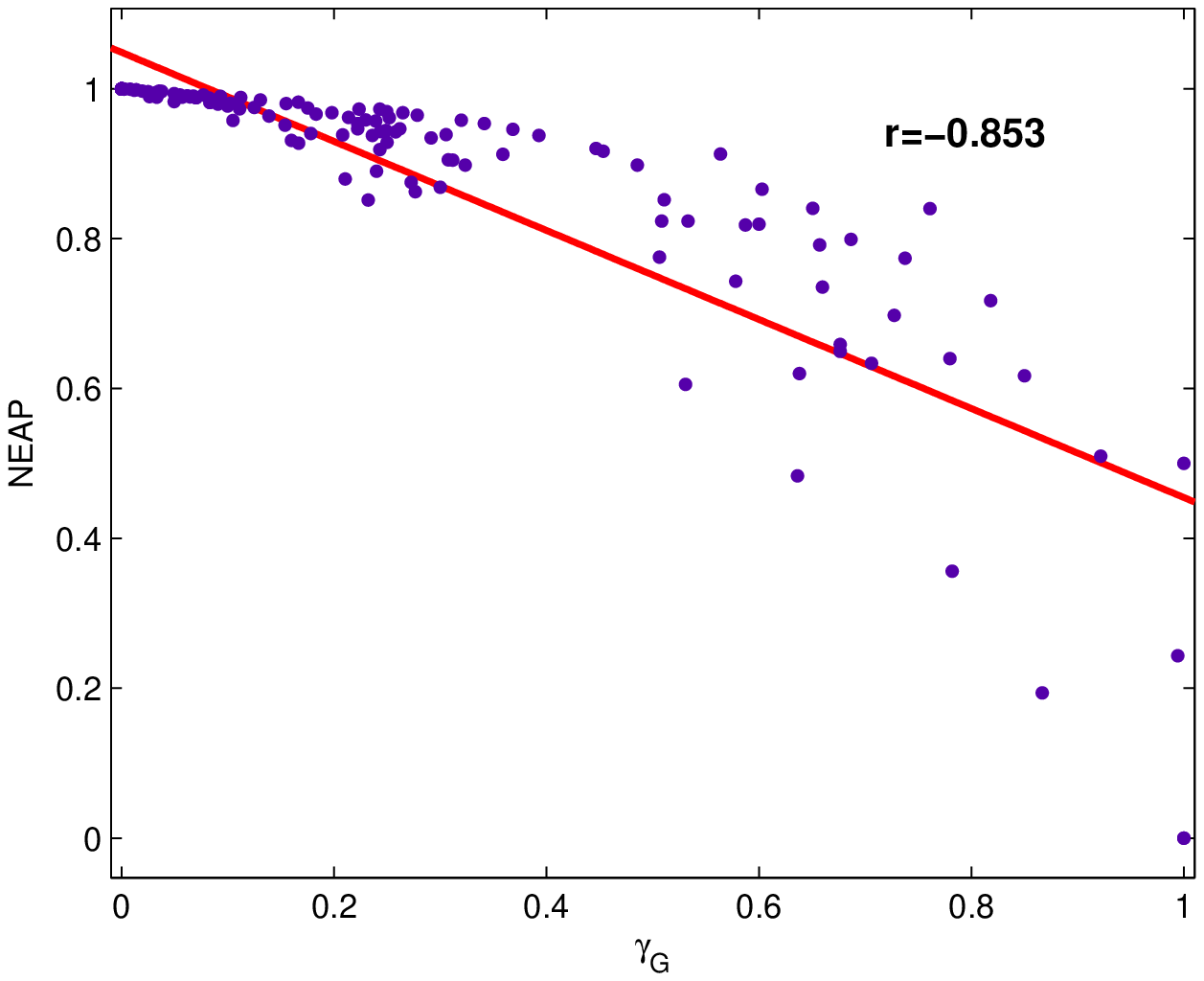}}
\caption{. $NEAP$ appears to be negatively correlated to $\beta_G$
and $\gamma_G$, and corresponding correlation coefficients are
-0.703 and -0.853, respectively. 125 Real networks and 28 theoretic
networks, overall 153 samples, are used.} \label{fig:correlation}
\end{figure}

Next, we will show that structural heterogeneity is strongly
negatively correlated to the symmetry of networks, which means that
the less symmetric a network is, the more structure-heterogenous the
networks is. As shown in Figure \ref{fig:correlation}, strong
negative relation could be observed from the $\beta_G-NEAP$ and
$\gamma_G-NEAP$ correlation curves. In fact, if a network is very
symmetric, nodes of the graph will have higher probability to be
equivalent in the structure, thus the automorphism partition will be
much closer to a \emph{unit partition}, which is extremely
homogenous. Conversely, if the network is closer to a asymmetric
network, vertex can be easily differentiated from each other from
the structural perspective, leading to a nearly discrete
automorphism partition. Consequently, the whole network tends to be
structure-heterogenous.
%

\section{\label{sec:level1}Conclusion}
We have shown that entropies based on degree partition can not
precisely describe the structural heterogeneity of complex networks
in many cases due to its inability to differentiate vertices with
the same degree. Instead, due to the strength of automorphism
partition that can naturally partition vertex set into equivalent
cells from the structure perspective, entropy based on automorphism
partition can quantify the heterogeneity of networks more
accurately.

Networks with extreme heterogeneity and homogeneity under different
entropy measures, including two degree-based entropy and symmetry
based entropy, have been analyzed, showing that symmetry-based
entropy is more accurate in quantifying the heterogeneity or
disorder of a network system than degree-based entropies. We also
calculated symmetry and heterogeneity statistics for hundreds of
real networks and several theoretic networks, and found that real
networks are more heterogenous in the view of automorphism partition
than what have been depicted under the measurement of degree-based
entropy. We also found that structural heterogeneity measured by
automorphism partition based-entropy is highly negatively correlated
to the abundance of symmetry in real networks.

Generally, heterogeneity of networks is strongly correlated to the
complexity of a network system, i.e., more heterogenous, more
complex. Thus, we believe that precisely characterizing the
heterogeneity of a network can definitely allow us to gain deeper
insight into the complexity of systems represented by network.

\section{\label{sec:level1}Acknowledgement}
\input acknowledgement_paragraph_r2.tex   

\section{\label{sec:level1}References}

\end{document}

%% file: list_of_authors_r2.tex
%
\author{Yanghua Xiao$^{1}$}
\author{Wentao Wu$^{1}$}
\author{Hui Wang$^{2}$}
\author{Momiao Xiong$^{3,4}$}
\author{Wei Wang$^{1}$}

\affiliation{\vspace{0.1 in}\vspace{0.1 in}}

\affiliation{$^{1}$Department of Computing and Information
Technology, Fudan University, ShangHai 200433, PR China}

\affiliation{$^{2}$Business school, University of Shanghai for
Science and Technology, Shanghai 200433, PR China}

\affiliation{$^{3}$Theoretical Systems Biology Lab , School of Life
Science, Fudan University, Shanghai 200093, PR China}

\affiliation{$^{4}$Human Genetics Center, University of Texas Health
Science Center at Houston, Houston TX 77225, USA}

%% file: acknowledgement_paragraph_r2.tex
The work was supported by the National Natural Science Foundation of
China under Grant No.60303008; the National Grand Fundamental
Research 973 Program of China under Grant No.2005CB321905